\documentclass{article}

\usepackage{authblk}
\usepackage{natbib}
\usepackage[letterpaper,top=2.5cm,bottom=2.5cm,left=3cm,right=3cm,marginparwidth=1.75cm]{geometry}

\usepackage{amsmath}
\usepackage[colorlinks=true, allcolors=blue]{hyperref}
\usepackage{graphicx}
\graphicspath{ {./figures/} }

\usepackage{footnote}
\usepackage{footmisc}

\usepackage{fancyhdr}
\usepackage{eso-pic}

\AddToShipoutPictureBG{
  \AtPageLowerLeft{%
    \raisebox{10pt}{\makebox[\paperwidth]{\begin{minipage}{21cm} \centering \color{gray}Preprint submitted to Handbook of Computational Social Science
    \end{minipage}}}%
  }
}

\title{Computational Approaches to the Study of Corruption\footnote{This is a draft chapter. The final version will be available in Handbook of Computational Social Science edited by Taha Yasseri, forthcoming 2023, Edward Elgar Publishing Ltd. The material cannot be used for any other purpose without further permission of the publisher, and is for private use only. Please cite as: Villamil, I., Kert\'esz, J., and Wachs, J. (2023). Computational Approaches to the Study of Corruption. In: T. Yasseri (Ed.), \textit{Handbook of Computational Social Science}. Edward Elgar Publishing Ltd.}}
\author[1]{Isabela Villamil}
\author[1,2]{J\'anos Kert\'esz}
\author[2,3]{Johannes Wachs}
\affil[1]{Central European University}
\affil[2]{Complexity Science Hub Vienna}
\affil[3]{Vienna University of Economics and Business}

\date{}

\begin{document}
\maketitle
\hrule
\begin{abstract}
Studying corruption presents unique challenges. Recent work in the spirit of computational social science exploits newly available data and methods to give a fresh perspective on this important topic. In this chapter we highlight some of these works, describing how they provide insights into classic social scientific questions about the structure and dynamics of corruption in society from micro to macro scales. We argue that corruption is fruitfully understood as a collective action problem that happens between embedded people and organizations. Computational methods like network science and agent-based modeling can give insights into such situations. We also present various (big) data sources that have been exploited to study corruption. We conclude by highlighting work in adjacent fields, for instance on the problems of collusion, tax evasion, organized crime, and the darkweb, and promising avenues for future work.
\end{abstract}

\vspace{16pt}
\hrule

\section{Introduction}

Corruption is an important and stubborn problem. It slows economic growth \citep{mauro1995corruption}, increases inequality \citep{gupta1998does}, and slows innovation \citep{rodriguez2014quality}. At the same time we know that high inequality fosters corruption \citep{jong2005comparative} suggesting that feedback loops between corruption and other social ills lead societies into vicious cycles. Corruption is thought to be a mechanism by which autocrats keep themselves in power \citep{fjelde2014political}, and it corrodes social trust \citep{rothstein2013corruption}.

Given its importance and prevalence, it is perhaps no surprise that many researchers study corruption. Yet despite these significant efforts, the research community around corruption increasingly laments the perceived failure of the anti-corruption research and policy agenda \citep{mungiu2017time,heywood2017rethinking}. Indeed corruption seems to be as much of a problem as ever. At the same time, the methods and ideas of computational social science are offering new ways to look at the phenomenon of corruption. By surveying these new lines of research here, we hope to highlight their potential to complement traditional approaches in the fight against corruption.

From a research perspective corruption presents several unique challenges, beginning with the question of how to define corruption. The World Bank and Transparency International define corruption as ``the abuse of public or corporate office for private gain'' and ``the abuse of entrusted power for private gain,'' respectively~\citep{nye1967corruption,transparency2007global}. A particularly influential framing of corruption is sometimes referred to as Klitgaard's formula: ``corruption equals monopoly plus discretion minus accountability'' \citep{klitgaard1988controlling}. These definitions have led anti-corruption policy actors to advocate for strong and independent public institutions to control corruption. Such institutions, for instance an ombudsman or independent anti-corruption agency, are supposed to limit abuses of power and increase accountability. Yet these approaches suffer from a chicken-or-egg style problem: there are many obstacles to setting up a working anti-corruption institution in an endemically corrupt environment \citep{heywood2017rethinking}. 

While defining corruption is difficult, measuring it consistently may be even harder. The most widely cited and discussed measures of corruption are based on national surveys in which people and prominent economic actors are asked about their perceptions of corruption. Perceptions of corruption can be misleading \citep{torsello2016anthropology}, repeating the survey data collection and measurement in a consistent manner is difficult \citep{heywood2014close}, and gathering subnational data is costly \citep{rothstein2013quality}. Other popular approaches to measure corruption have drawbacks too. Large scale randomized experiments provide invaluable evidence though are often prohibitively expensive and necessarily focus on highly particular questions \citep{olken2007monitoring,bertrand2007obtaining,ferraz2008exposing}. Lab experiments make relevant contributions to our understanding of the psychological and perhaps even cultural roots of corruption \citep{weisel2015collaborative}, but are limited in scale and generalizability. Observational data about corrupt behavior is difficult to come by and is often biased: one may not want to learn too much from the crooks that were caught.

More recent theoretical work has recognized corruption rather as a collective action problem, in which it is reasonable to expect individuals to participate in corruption if many are doing so in a society \citep{persson2013anticorruption}. This perspective suggests how corruption is a relational phenomenon, and that actors are embedded in environments with varying levels of corruption \citep{granovetter1985economic}. Indeed, evidence suggests that corruption is both  contagious \citep{gulino2021contagious} and collaborative \citep{weisel2015collaborative}. Corruption seems to be more than the sum of a series of bribes or favors exchanged; it is rather a collection of ways to organize the extraction of economic and political rents from the state \citep{grzymala2008beyond}. While this conceptualization of corruption as a complex, networked phenomenon is not a new idea \citep{della2016hidden}, the new kinds of data and methods typical of computational social science are ideally suited to investigate corruption from this perspective \citep{kertesz2021complexity}. In other words, computational methods and data sources have developed so that we can better test and build collective action theories of corruption.

In this chapter we highlight three recently flourishing areas of corruption research with a strong computational social science flavor. The first seeks to map the structure and dynamics of large scale corruption and related activities using diverse data sources. The second area uses mines public procurement or purchasing data for signals of corruption risk. We also highlight how agent-based models are used to explore the emergent and mesoscopic properties of corruption. We conclude the chapter by highlighting especially promising areas for future work.

\section{Computational Social Science in Corruption Research}

\subsection{The Structure and Dynamics of Corruption Networks}

The amount of data that has recently become available to researchers has provided an opportunity for richer studies to help us understand corruption. However, traditional tools of analysis have proven to be insufficient given the volume of information as well as the increasing complexity of relationships between actors involved in corruption schemes. Computational techniques have become valuable to characterize the structure and dynamics of corruption networks. Other studies examining the robustness and vulnerabilities of complex networks can assist in finding ways to efficiently disrupt the activity of corrupt individuals and organizations.

\begin{figure}[h]
\centering
\includegraphics[width=0.6\textwidth]{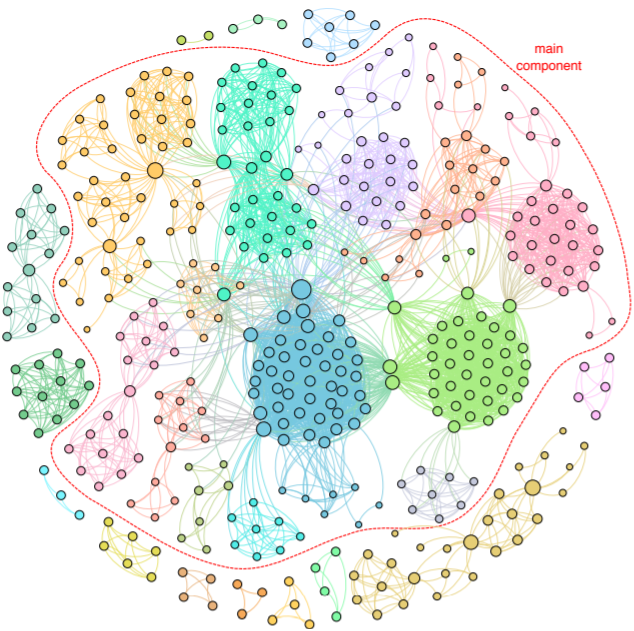}
\caption{Network of individuals involved in major corruption cases in Brazil from 1987 to 2014, as studied in Ribeiro et al. \citeyearpar{ribeiro2018dynamical}. Nodes represent people and are connected by an edge when the two individuals appear at least once in the same corruption scandal. Node sizes are proportional to their degrees while node color indicates communities obtained using the network cartography approach. The structure of the network is indicative of tight links between different scandals, with participants having intricate relationships with one another.}
\label{figure:brazil}
\end{figure}

\cite{ribeiro2018dynamical} use data on convicted public officials extracted from Wikipedia to examine the emergence of systemic corruption in Brazil. Covering 65 major corruption cases that occurred over 27 years, the authors found that the co-occurrence of politicians in corruption scandals create networks with connected components that can span decades. By the time the Petrobras scandal erupted in 2014, a giant connected component had emerged in the network of co-conspiracy, connecting a large majority of individuals (see Figure \ref{figure:brazil}). The same dataset was used by \cite{ren2019generalized} in their paper proposing a framework for cost-efficient network dismantling. A typical approach to fight organized crime is to try to identify the underlying network of actors involved in the conspiracy and then to remove the ringleader. However, the authors argue that a more effective strategy to dismantle a corruption network is to remove medium-sized nodes. Another work that uses a network-based approach to study corruption is by \cite{colliri2019analyzing}. The authors used data from almost 30 years of bill-voting in Brazil to examine the relationship between voting history and convictions of corruption or other financial crimes among legislators. They found the formation of so-called “corruption neighborhoods” in the network, indicating that corrupt officials tend to have similar voting patterns. Using link prediction techniques, the authors were able to successfully predict the involvement of lawmakers in criminal activity.

A new kind of data that has emerged as a promising source for case studies are the series of large-scale document leaks involving offshore service providers. One of the most influential of these is the so-called Panama Papers, released by the International Consortium of Investigative Journalists (ICIJ) in 2016. Until recently, the Panama Papers was the largest ever leak of offshore documents. These papers represent an important milestone in the use of data science in investigative journalism. The difficultly posed by the leak was not just in its size, but the fact that its importance came from the relationships it established between specific entities and individuals. To overcome this issue, the ICIJ used tools from data mining and network science to identify and map these links, finding criminal and corruption connections. 

So far, academic studies using computational techniques that leverage these offshore leaks have been limited. A few describe the structure and properties of the offshore network to identify important actors and groups \citep{kejriwal2020structural,dominguez2020panama}. One possible reason for the dearth in research is the constraint imposed of having information derived from a limited number of sources. Offshore leaks are often obtained from a few large firms and provide an incomplete picture of the actors and links involved in corruption schemes. However, the increasing number of leaks, such as the Pandora Papers released in 2021, can provide us with even more information on illegal offshore activities. Combining these databases with other sources also allows for deeper analysis. \cite{joaristi2019social} use known lists of suspicious entities to verify ground truth of criminal activity, proposing a new algorithm inspired by PageRank to identify suspicious companies and people that are more likely to be involved in illegal activities.

One of the main purposes of using offshore companies in criminal activity is to hide ownership structures. Studies on corporate ownership can therefore help untangle the complex relationships between firms to determine the true beneficial owners. At the same time, there is evidence that firms are connected beyond common shareholders. \cite{jeude2019multilayer} construct a multiplex network of interactions between companies in Germany and in the United Kingdom, combining ownership links, interlocking directorates, research and development collaborations, and stock correlations. They found a non-trivial overlap between these different types of networks, where the different types of connections complement each other and make the overall structure more complex. This highlights that corporate control, boardroom influence and other connections have different structures and together make an even smaller corporate world than previously thought.

The offshore leaks expose two important observations in contemporary corruption cases. First, is the transnational nature of large-scale corruption. While offshore companies are not in themselves illegal, they can facilitate the conduct of illegal practices. For instance, in a forthcoming paper, \cite{andersen2021elite} discovered a link between aid payments to developing countries and in-flows into financial centers known for bank secrecy and private wealth management. The pattern of flows suggests that a portion of the aid received by corrupt countries are diverted to the private accounts of ruling politicians and bureaucrats. Identifying offshore financial centers has therefore become a critical, albeit sensitive, part of efforts to understand and control corruption. \cite{garcia2017uncovering} propose the use of network science methods to objectively classify such jurisdictions. Using data on global corporate ownership, the authors highlight jurisdictions that serve as so-called sinks and conduits of the tax-offshoring industry. They found that although a great deal of money ends up in the sinks, which are the typical island tax havens, it turns out that the conduits through which money flows on the way to the sinks are often larger countries, including the Netherlands and the UK. Their results call attention to the limitations of the more popular perception-based measures of corruption, which often show that corruption is disproportionately a problem of the developing world. Such measures disregard that a sizeable proportion of large-scale corruption is conducted via transnational networks involving intermediaries in richer countries who provide options for conspicuous consumption \citep{cooley2018how}. 

A second important observation from the offshore leaks is the use of shell companies for concealing corrupt acts and proceeds. The Panama Papers hinted at the extent to which public figures use offshore shell companies in order to hide their real wealth. However, corrupt policy makers have also used domestic shell companies that function as intermediaries to divert public resources for private gain. One study \citep{luna2020corruption} describes a major corruption scandal that took place in Mexico involving a network of hundreds of shell companies that were awarded contracts for public projects that were never delivered. Using a dataset collected from official sources, the authors generate a bipartite multigraph linking companies and individuals through various type of business relationships. They found that ownership links provide the most cohesion to the network. However,  the shell companies were also connected through entities performing other roles (e.g., legal representatives and administrators). Interestingly, aggregating all layers of relationships resulted in a fully connected network. This shows the complex interrelationships that can be formed between actors performing various roles, whether licit or illicit, in a corruption scheme.

As illustrated by the scandals in Mexico as well as other countries, government procurement is an activity vulnerable to corruption. The procurement process provides opportunities for various forms of corruption, such as embezzlement, graft, bribery, cronyism, and fraud. With the development of computational techniques, there has been a growing academic literature using more sophisticated analyses to study corruption in public procurement. Some of these studies are discussed in more detail in the next section.

\subsection{Measuring Corruption via Procurement Data}
While analysis of specific cases provides valuable insight, this approach cannot lead to a general, explanatory picture of how corruption works. For example, it is likely that more effective corruption rings survive longer without detection. Even if all corrupt groups operated the same way, it is often unclear how to apply the lessons learned from their specific situation to search the wider world for similar ones. A recent line of research takes an entirely different and in some sense the opposite approach. These papers study large datasets of public procurement contracts and search for indicators of corruption risk. Rather than attempting to generalize from verified cases, this approach searches for signals of corruption in complete or at least very large databases of contracts.

Public procurement is the process by which the public sector purchases goods and services from the private sector. It accounts for a significant share of the global economy, with estimates from OECD countries ranging from 5-20\% of national GDP \citep{oecdprocurement}. While procurement is by no means the only lever of the state that can be used to favor individuals and firms, it is perhaps the largest direct channel of funds from public to private. Procurement also has widely accepted norms and best-practices: free and fair competition for contracts is thought to deliver the best value for money. Typically, procurement contracts are awarded to the firm that has submitted the bid satisfying the conditions with the lowest price offer. Finally, governments are required to publish data on public procurement and tend to do so.

With such data in hand, researchers search for deviations from norms of impartiality and best practices \citep{fazekas2020uncovering}. For instance, a complex contract first advertised just three days before the bid submission deadline may indicate that an insider has tipped off a favored firm. Likewise, undue restrictions to participate, for example that a firm has to have highly particular previous experience, can all but insure that competition is limited. These and other similar red flags or corruption risk indicators, as they are referred to in the literature, predict low levels of competition and high prices \citep{fazekas2020uncovering}. In aggregate, the share of procurement awarded as single-bid contracts is significantly correlated with national \citep{wachs2021corruption} and regional measures \citep{fazekas2021measuring} of corruption drawn from surveys or measures of institutions.

Granting that no amount of red flags on a contract can prove that it was awarded in a corrupt procedure, 
these procurement indicators of corruption risk are useful because they can be aggregated across various scales and times to revisit and test old and often disputed questions about the nature of corruption and its effects. For example, political incumbency and entrenchment are thought to predict greater levels of corruption. Broms et al. \citeyearpar{broms2019political} found that Swedish cities with a long incumbent mayor issues significantly more single-bid contracts. Similarly, a mayor's previous larger victory margin predicts greater single bidding rates in Hungarian towns \citep{wachs2019social}. Procurement-based corruption proxies enable researchers to compare hundreds of cities and towns within the same country.

The fine-grained nature of procurement data also allows researchers to study particular policy alternatives. In a study of European regions, Charron et al. found that corruption risks in procurement are substantially lower where bureaucrats have meritocratic career paths \citeyearpar{charron2017careers}. Perhaps unsurprisingly, research also suggests that political campaign contributions are a vehicle for favoritism: in the US firms donating to political parties tend to earn more procurement revenue, and do so by winning contracts with greater corruption risk \citep{witko2011campaign,fazekas2021agency}.

The response of procurement markets to electoral turnover and volatility also indicates potential corruption risk. One study shows a marked spike in first time winners and surprising losers in contracting markets following changes in government both in the UK and in Hungary \citep{david2020grand}. While about 10\% of firms in the UK fall into these potentially politically connected categories, in Hungary the estimate is closer to 50\%. Another study looked at the bipartite networks of contract issuers and winners in Hungary and the Czech Republic, finding that the network neighborhoods of issuers of high corruption risk contracts tend to experience much greater volatility when the government changes \citep{fazekas2020corruption}.

The mapping of contracting markets as networks also allows us to study the distribution of corruption risks within a country or region. In a study of European markets, we found that procurement markets have highly heterogeneous structure \citep{wachs2021corruption}. As is typically the case with empirical networks, the typical market participant wins or issues just a few contracts, while the largest hubs win or issue hundreds in a year. These networks are also modular and have high levels of clustering. They tend to vary in two important ways relevant to the study of corruption risk. First, the distribution of high risk (e.g., single-bid) contracts is highly non-uniform in all countries studied. Second, the degree of centralization of the market and the tendency for high risk contracts to appear in the core or periphery of a market vary from country to country.

\begin{figure}[h]
\centering
\includegraphics[width=0.8\textwidth]{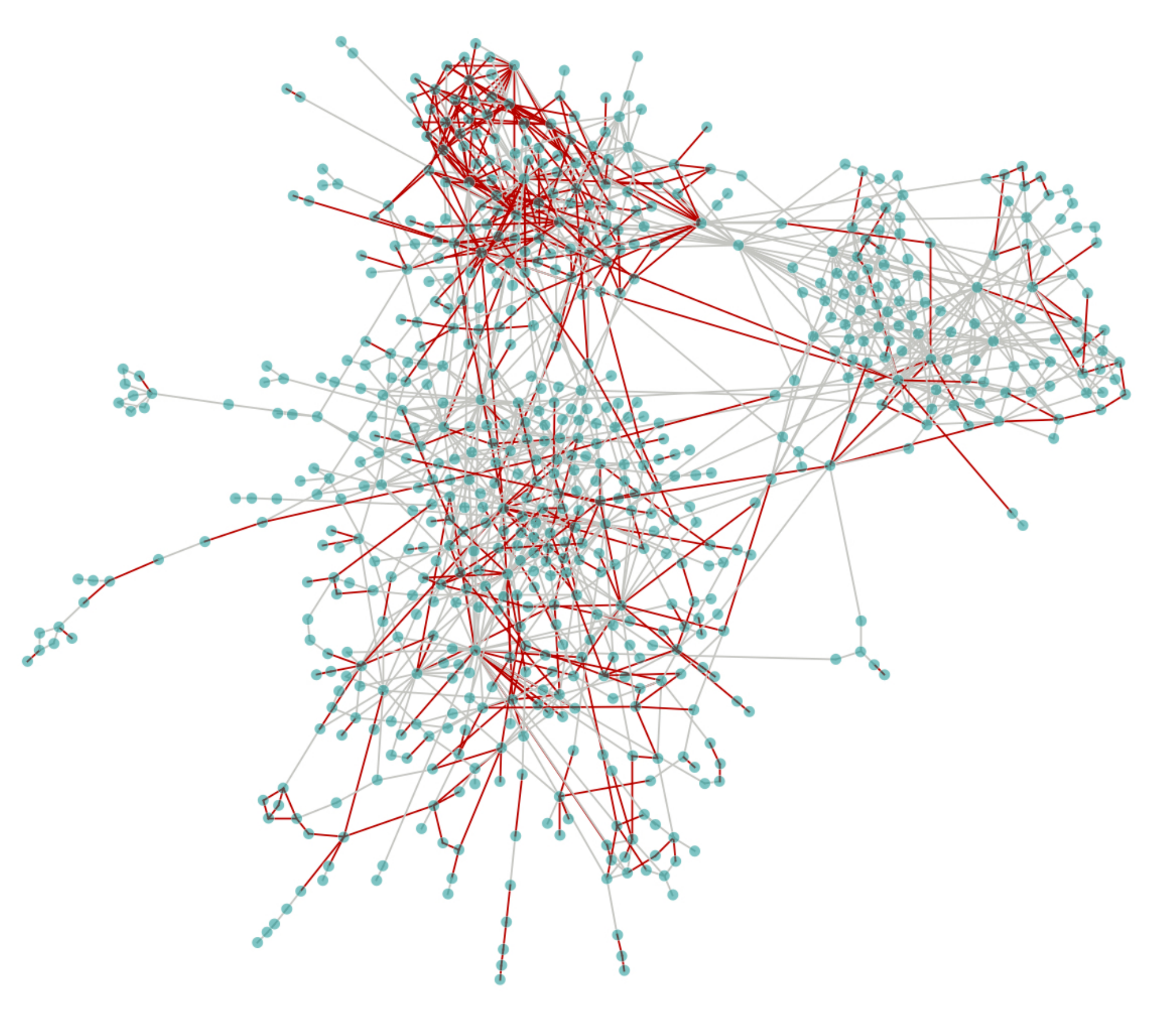}
\caption{The largest connected component of the 2014 Hungarian procurement market represented as a network, from Wachs et al. \citeyearpar{wachs2021corruption}. Nodes are buyers and suppliers of contracts, connected by an edge if they contract with one another. Edges are colored red if the single-bidding rate (a key marker of corruption risk) on the edge exceeds the average rate of single bidding that year. Single bidding is significantly overrepresented among the edges in the top left cluster.
}
\label{figure:procurement_net}
\end{figure}

The first observation suggests that there is significant heterogeneity in the level of corruption risk in different parts of a country, whether defined by sector of the economy or public administration, or geography. Yet the degree of clustering of corruption risk is also found to be robust to randomization within sector, suggesting that there are idiosyncratic high risk parts of the market that can be investigated and studied. We visualize one year of procurement data from Hungary as a network in Figure~\ref{figure:procurement_net}, highlighting contracting relationships with high amounts of single-bidding in red.

The distribution of high risk contracts across the core and periphery of the market links this work to an ongoing debate in political science and economics. Researchers have long disputed whether centralized government and public administration hinders or fosters corruption. While a highly centralized government may obscure political responsibility for corrupt outcomes and weaken the link between local outcomes and political feedback \cite{persson2002political}, previous work finds no significant correlation between centralization and corruption measured via surveys \cite{charron2014regional}. We found that countries with greater corruption risk on average, tended to have more contracts in the core of the market, i.e. greater centralization (Spearman's correlation $\rho>0.64$). At the same time, the centralization of contracts in general - a purely network-topological observation about a country's procurement market - had a significant negative correlation with corruption as measured by Charron et al. \citeyearpar{charron2014regional}. In general the topology of procurement markets and the distribution of corruption risk within them seems a fruitful way to understand the organizational modes of corruption, linking naturally to concepts such as state capture \citep{fazekas2016corruption} and organized market manipulation \citep{lyra021firmfirm}. However, care must be taken to study these networks using appropriate methods, as they vary significantly in size \citep{curado2021scaling}.

We finally discuss a work using procurement data to proxy for corruption risk which zooms in on local social conditions and their relationship to corruption outcomes. Using data from a once widely used online social network, we related the level of corruption risk in the procurement procedures of Hungarian towns with their social network structure \citep{wachs2019social}. Drawing on the theory of social capital, we found that towns with a high level bonding social capital tended to have higher corruption risk in their contracts. Bonding social capital is high when individuals have tightly linked networks of social contacts. Aggregated to the level of a town or city, such networks appear highly fragmented, as individuals split into communities. We theorized that corruption manifests as the end result of in-group favoritism; this is facilitated when a place has clearly de-lineated groups. Bridging social capital, the tendency for individuals to have diverse connections, including beyond the borders of their hometown, was a strong predictor of lower corruption risk. These findings provide additional evidence that corruption is an embedded phenomenon in places, and suggests why corruption tends to be so persistent in specific places: social networks are slow to change over time.

\subsection{Agent-Based Models}
We have discussed at some length that corruption is a social phenomenon, with clear characteristics of complexity: many interacting players, adaptivity, heterogeneity, non-linearity, feedback mechanisms, etc. Furthermore, features like different moral attitudes, heterogeneous risk taking, deceptive behavior, frequent involvement of those who are supposed to fight corruption make its systematic scientific treatment particularly difficult. And while theoretical and qualitative work has generated important insight into the nature of corruption, it seems difficult to put forward theories on quantitative bases. Big Data, which has its impact on this field too, has mostly been used for case studies and to measure corruption risks, as discussed in the previous sections. Occasionally, observational studies have lead to the identification of interesting relationships between social features and the level of corruption risk \citep{wachs2019social} and some attempts have been made towards working out a typology of the procurement markets as bipartite networks of issuers and winners of contracts, as discussed in the previous section.

Nevertheless, we are very far from a comprehensive quantitative description - in fact, we have not even had a real start on it. Mean field or "representative agent" type of theories have been put forward (see, e.g.,~\cite{Olken2012Developing}), which are able to capture some relationships between variables like wage of the bribed person, the bribe, the dishonesty costs and the probability of being detected. However, such equilibrium theories ignore fluctuations and the concept of bounded rationality which are essential ingredients of the systems under consideration. 

A rather new and promising methodological approach to complex systems is Agent-Based Modeling (ABM~\citep{Gilbert2008ABM,Railsback2019ABMbook}). In ABM the system can be described down to the individual actors, who are assumed to be heterogeneous both in their features and interactions. The interactions take place on a complex network and the processes take place on it. An agent-based model is defined by setting the agents, their strategies, the interactions, as well as the environment. In an idealized case this flexible framework should include all players and the full variability of their features and interactions together with the economic, legal, and political environment. Such a complex system would have a self-evolving dynamics producing emergent phenomena, including corruption. There are, however, several problems with this construction, most importantly, we do not know much of the ingredients which define the model, nevermind the computational limitations. Therefore a different strategy is followed in the spirit of how science proceeds: We try to focus on the relevant aspects of the problem and identify the main parameters learning this way about the leading mechanisms.

A natural starting point is game theory \citep{Osborne2010Gametheory} as decisions related to corruption contain much of the situations grabbed by basic strategic games. This was recognized already in the 80's \citep{MacRae1982Underdevelopment}, when it was shown that rather natural assumptions about the relationship between bribe, price, and the probability to win the contract under the uncertainty about the behavior of the competitors create decision situations very similar to that of the prisoner's game, one of the basic game theoretical models. These models treat corruption as isolated, individual decision problems. The social aspect can be pursued in the more elaborated setting of ABM \citep{Elnawawy2019ABM_review}.

In his early model Hammond ~\citeyearpar{Hammond2000Endogenous} already set up a complex machinery: members of two groups of interacting agents (bureaucrats and citizens) form friendship network and interact at every instant of time with randomly chosen partners of the other group by choosing either corrupt or honest strategy. Local information is available for them and they have a finite memory. A payoff matrix forms the basis of the decision about the strategy as it is usual in game theory, however, a large amount of heterogeneity is introduced by the diversity of the predisposition of the agents toward corrupt action and the random interactions. There is a parameter governing the risk for a corrupt agent to be put into jail for some time. The main finding of the paper is that given the parameter settings an initially corrupt and seemingly stationary state of the system suddenly changes to be predominantly honest without any external effect. This observation has an optimistic message, though is unfortunately not in agreement with empirical observations, but the model directs the attention to the importance of endogenous processes. 

\begin{figure}[h]
\centering
\includegraphics[width=0.8\textwidth]{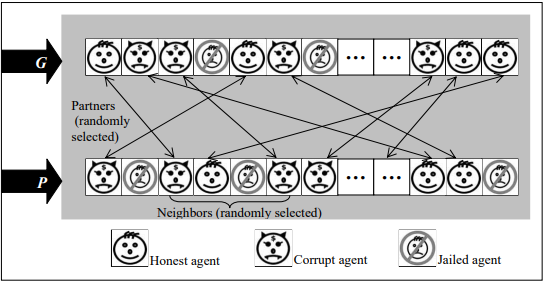}
\caption{The scheme of the ABM from \citep{Situngkir2004StructuralDyn}: There are two groups of agents in the model, G and P. Within a group random environments are defined from which information can be gathered. Randomly selected pairs from the two groups interact with each other according to their individual characteristics and a payoff matrix resulting in honest or corrupt behavior.}
\label{figure:ABM}
\end{figure}

\cite{Situngkir2003Money-Scape} used an iterated game-theory
approach as a starting point, where the agents (bureaucrats) were able to learn by watching how many of their neighbors on a square grid end up in jail. This model also includes heterogeneous levels of honesty and risk aversion for the agents as well as the benefits of both corrupt and honest acts. In the spirit of evolutionary games \citep{Szabo2007Evolutionary}, agents learn during the course of the simulation and improve their behaviors. This process leads to a kind of (Nash) equilibrium, where no agent can improve its situation. The conclusion of this study is just the opposite of that of~\cite{Hammond2000Endogenous}: The system often converges toward an overall corrupt state. Later the same author introduced a more sophisticated model \citep{Situngkir2004StructuralDyn} with three types of agents (citizens, bureaucrats, law-makers) and demonstrated in his model that political corruption has the consequence that citizens become corrupt. For the schematic structure of the model see Figure ~\ref{figure:ABM}.

The ABM by~\cite{Chakrabarti2007Chapter} showed that besides risk aversion human capital plays a role. The overall corruption level results from the interaction of the agents who optimize their behavior according to their different features. This is a self-consistent process, as the societal corruption level reacts to the individual behavior. The surprising result is that there are stable solutions with different corruption levels depending on the socio-economic parameters and, in some cases, the corruption can grow to an extent that it practically paralyzes the economy.

In the recent paper by~\cite{Zausinova2020Complexity} several aspects of previous models are combined. The goal was to demonstrate that corruption is a phenomenon of the society, where a number of effects studied in complexity science occur, such as adaptivity, non-linearity, self-organization and sensitivity to the initial conditions. In this versatile model one can take into account the historical-cultural embedding, the legal environment, and the role of uncertainty and fluctuations. Accordingly, a diversity of phenomena could be observed like the emergence of corruption and the  convergence toward a corruption free equilibrium under some conditions.

A further recent remarkable paper studied the network effects on corruption \citep{ferrali2020partners}. Inside the organizational network structure corrupt agents form a criminal subnetwork with its own recruitment and security rules. An important finding is that isolated parts of the original organizational network carry enhanced corruption risk. Interestingly, changing the network structure (enhancing its capacity) by eliminating such isolated parts helps fight "petty corruption" but "grand corruption" remains, involving more accomplices. This paper used methods of theoretical considerations, ABM and real experiments, with the results supporting one another.

Here we do not have the opportunity to present the entire literature on ABM of corruption -- we refer the interested reader to a recent more comprehensive review by~\cite{Elnawawy2019ABM_review}. Instead, we would like to summarize the pros and cons of this method and present our opinion about its perspectives.  In 2009 \citep{Farmer2009Nature} argued that "The economy needs ABM". As they write: "Agent-based models potentially present a way to model the financial economy as a complex system, ... while taking human adaptation and learning into account... Such models allow for the creation of a kind of virtual universe, in which many players can act in complex — and realistic — ways." This is even more true for economic crime, where not only the inherent complexity of the economy is present but, in addition, cultural, historical, and psycho-sociological components. We believe therefore that there is great potential in applying ABM to studying corruption. Already the studies so far have revealed important features related to corruption and explored their impact and interplay. Models include relevant factors as level of honesty, level of risk aversion, local versus global information, inhomogeneity in these components, role of penalty, etc. 

With all these attractive features it should be noted that ABM has its teething problems. There is no generally accepted methodology, the rules of calibrating and verifying the models are not set; in general, relating the computer experiments to observations is a major challenge. The difficulties are particularly present for the application of ABM to corruption. No wonder that most of the related papers have been published in less established or lower impact journals (if at all). Nevertheless, given the complexity, social roots, and hidden nature of corruption, the ever evolving ABM seems indispensable in the toolbox of its scientific approaches.

\section{Conclusion}

To conclude we reflect on the future potential of computational social science research on corruption. We highlight the impact of computational methods on neighboring fields, mention promising emerging datasets and sources, and speculate on the potential of un- or underexplored ideas. Finally, we reflect on the cat-and-mouse nature of anti-corruption research, noting that corrupt actors are also learning and making use of new technologies. 

In recent years the use of computational methods has greatly expanded in a the broader context of criminology \citep{campedelli2021we,campedellimachine}. New data sources are being used to study traditional organizational forms of criminal behavior, such as the mafia \citep{cavallaro2020disrupting} and gangs \cite{papachristos2013corner}. The methods of network science have also been applied to study police misconduct \cite{wood2019network}. Methods such as link prediction and centrality analysis can provide insights into the key players of various conspiracies \citep{berlusconi2016link}. Agent-based models are used to evaluate policing strategies \citep{weisburd2017can}. More broadly, however, they can describe the structural and mesoscopic aspects of organized criminal activity. Such work often highlights connections between criminal worlds and systems, for instance between mafia activity and political corruption \citep{joseph2021ties}. The application of computational methods in the allocation of law enforcement resources is a topic of significant controversy, given that these systems are often found to encode and even reinforce biases \citep{lum2016predict,ensign2018runaway}.

At the same time, data on economic activity and transactions, especially adjacent to the public domain, are becoming increasingly well-organized and available. Governments around the world are increasingly adopting the Open Contracting Data Standard \citep{corcho2019towards}, which aims to standardize data on procurement awards. In most procurement portals, the disambiguation and deduplication of actors and the estimation of contract value, to take two examples, present significant hurdles to quantitative research on procurement. Another effort called Open Corporates has the goal of publishing comparable information on firms \citep{soylu2019overview}. On the other hand, data on beneficial ownership links of companies, especially cross-border links, remains limited despite pressure on governments from advocacy groups \citep{kurniawan2020disclosure}. 

In addition to the proliferation of datasets and the development of new methods, the web itself is changing how we can study corruption. Public procurement portals like the Ukranian Prozorro allows citizens to comment and discuss contract awards in real time \citep{nizhnikau2020love}. In Hungary, citizens have organized the ``K-Monitor'' platform, on which individuals can report suspicions of corruption, organize collaborative investigations and anti-corruption campaigns \citep{pirro2021corruption, jancsics2017offshoring}. Both are examples of crowdsourced anti-corruption activism tools which are made possible by the web. Social media too can foster anti-corruption activity in places where media and political competition are limited \citep{enikolopov2018social}. Yet we know little about how social media influences political attitudes about corruption, especially when it is used as an organizational tool by activists.

Despite the potential and promise of emerging ICT tools like e-government services, transparency portals, whistleblowing tools, and distributed ledgers \citep{bertot2010using}, more work is needed to show how these technologies can moderate corruption, for instance by improving transparency and advocacy \citep{adam2021emerging}. Secure and anonymous communication channels and dark web markets can also facilitate corruption. Investigations of data from darknet markets \citep{duxbury2018network} and bitcoin ransomware attacks \citep{paquet2019ransomware} indicate how digitalization has expanded the scale and scope of criminal activity. More generally, the complexity of the digital world can just as easily be a tool for corruption as a cure. We believe that the perspective of computational social science and its practitioners will only become more valuable in the field of anti-corruption because of these changes.

\section{Further readings}
We share references on the topic that are particularly relevant:
\begin{itemize}
    \item A rallying call for evidence-based anti-corruption: \citep{mungiu2017time}
    \item A cross-national dataset of public procurement contracts rated for corruption risk: \citep{fazekas2020uncovering}
    \item Procurement markets mapped as networks: \citep{wachs2021corruption}.
    \item A review of agent-based models in the anti-corruption literature:~\cite{Elnawawy2019ABM_review}, 
    \item On emerging ICT and impacts on good governance: \citep{adam2021emerging}
\end{itemize}

\bibliography{main.bib}

\begin{thebibliography}{86}
\providecommand{\natexlab}[1]{#1}
\providecommand{\url}[1]{\texttt{#1}}
\expandafter\ifx\csname urlstyle\endcsname\relax
  \providecommand{\doi}[1]{doi: #1}\else
  \providecommand{\doi}{doi: \begingroup \urlstyle{rm}\Url}\fi

\bibitem[Adam and Fazekas(2021)]{adam2021emerging}
Isabelle Adam and Mih{\'a}ly Fazekas.
\newblock Are emerging technologies helping win the fight against corruption? a
  review of the state of evidence.
\newblock \emph{Information Economics and Policy}, page 100950, 2021.

\bibitem[Andersen et~al.(2021)Andersen, Johannesen, and
  Rijkers]{andersen2021elite}
J{\o}rgen~Juel Andersen, Niels Johannesen, and Bob Rijkers.
\newblock Elite capture of foreign aid: Evidence from offshore bank accounts.
\newblock \emph{Journal of Political Economy}, 2021.

\bibitem[Berlusconi et~al.(2016)Berlusconi, Calderoni, Parolini, Verani, and
  Piccardi]{berlusconi2016link}
Giulia Berlusconi, Francesco Calderoni, Nicola Parolini, Marco Verani, and
  Carlo Piccardi.
\newblock Link prediction in criminal networks: A tool for criminal
  intelligence analysis.
\newblock \emph{PloS one}, 11\penalty0 (4):\penalty0 e0154244, 2016.

\bibitem[Bertot et~al.(2010)Bertot, Jaeger, and Grimes]{bertot2010using}
John~C Bertot, Paul~T Jaeger, and Justin~M Grimes.
\newblock Using icts to create a culture of transparency: E-government and
  social media as openness and anti-corruption tools for societies.
\newblock \emph{Government information quarterly}, 27\penalty0 (3):\penalty0
  264--271, 2010.

\bibitem[Bertrand et~al.(2007)Bertrand, Djankov, Hanna, and
  Mullainathan]{bertrand2007obtaining}
Marianne Bertrand, Simeon Djankov, Rema Hanna, and Sendhil Mullainathan.
\newblock Obtaining a driver's license in india: an experimental approach to
  studying corruption.
\newblock \emph{The Quarterly Journal of Economics}, 122\penalty0 (4):\penalty0
  1639--1676, 2007.

\bibitem[Broms et~al.(2019)Broms, Dahlstr{\"o}m, and
  Fazekas]{broms2019political}
Rasmus Broms, Carl Dahlstr{\"o}m, and Mih{\'a}ly Fazekas.
\newblock Political competition and public procurement outcomes.
\newblock \emph{Comparative Political Studies}, 52\penalty0 (9):\penalty0
  1259--1292, 2019.

\bibitem[Campedelli(2021)]{campedelli2021we}
Gian~Maria Campedelli.
\newblock Where are we? using scopus to map the literature at the intersection
  between artificial intelligence and research on crime.
\newblock \emph{Journal of Computational Social Science}, 4\penalty0
  (2):\penalty0 503--530, 2021.

\bibitem[Campedelli(2022)]{campedellimachine}
Gian~Maria Campedelli.
\newblock Machine learning for criminology and crime research: At the
  crossroads.
\newblock 2022.

\bibitem[Cavallaro et~al.(2020)Cavallaro, Ficara, De~Meo, Fiumara, Catanese,
  Bagdasar, Song, and Liotta]{cavallaro2020disrupting}
Lucia Cavallaro, Annamaria Ficara, Pasquale De~Meo, Giacomo Fiumara, Salvatore
  Catanese, Ovidiu Bagdasar, Wei Song, and Antonio Liotta.
\newblock Disrupting resilient criminal networks through data analysis: The
  case of sicilian mafia.
\newblock \emph{Plos one}, 15\penalty0 (8):\penalty0 e0236476, 2020.

\bibitem[Chakrabarti(2007)]{Chakrabarti2007Chapter}
Rajesh Chakrabarti.
\newblock A dynamic agent-based model of corruption.
\newblock In \emph{Handbook of Research on Nature-Inspired Computing for
  Economics and Management}, chapter~9, pages 111--122. IGI Global, 2007.

\bibitem[Charron et~al.(2014)Charron, Dijkstra, and
  Lapuente]{charron2014regional}
Nicholas Charron, Lewis Dijkstra, and Victor Lapuente.
\newblock Regional governance matters: Quality of government within european
  union member states.
\newblock \emph{Regional Studies}, 48\penalty0 (1):\penalty0 68--90, 2014.

\bibitem[Charron et~al.(2017)Charron, Dahlstr{\"o}m, Fazekas, and
  Lapuente]{charron2017careers}
Nicholas Charron, Carl Dahlstr{\"o}m, Mihaly Fazekas, and Victor Lapuente.
\newblock Careers, connections, and corruption risks: Investigating the impact
  of bureaucratic meritocracy on public procurement processes.
\newblock \emph{The Journal of Politics}, 79\penalty0 (1):\penalty0 89--104,
  2017.

\bibitem[Colliri and Zhao(2019)]{colliri2019analyzing}
Tiago Colliri and Liang Zhao.
\newblock Analyzing the bills-voting dynamics and predicting
  corruption-convictions among brazilian congressmen through temporal networks.
\newblock \emph{Scientific Reports}, 9, 2019.

\bibitem[Cooley(2018)]{cooley2018how}
Alexander Cooley.
\newblock \emph{How International Rankings Constitute and Limit Our
  Understanding of Global Governance Challenges: The Case of Corruption}, pages
  49--67.
\newblock Springer International Publishing, 2018.

\bibitem[Corcho et~al.(2019)Corcho, Simperl, Konstantinidis, and
  Lech]{corcho2019towards}
Oscar Corcho, Elena Simperl, George Konstantinidis, and Till~Christopher Lech.
\newblock Towards an ontology for public procurement based on the open
  contracting data standard.
\newblock \emph{Digital Transformation for a Sustainable Society in the 21st
  Century}, page 230, 2019.

\bibitem[Curado et~al.(2021)Curado, Dam{\'a}sio, Encarna{\c{c}}{\~a}o, Candia,
  and Pinheiro]{curado2021scaling}
Ant{\'o}nio Curado, Bruno Dam{\'a}sio, Sara Encarna{\c{c}}{\~a}o, Cristian
  Candia, and Fl{\'a}vio~L Pinheiro.
\newblock Scaling behavior of public procurement activity.
\newblock \emph{PloS one}, 16\penalty0 (12):\penalty0 e0260806, 2021.

\bibitem[D{\'a}vid-Barrett and Fazekas(2020)]{david2020grand}
Elizabeth D{\'a}vid-Barrett and Mih{\'a}ly Fazekas.
\newblock Grand corruption and government change: An analysis of partisan
  favoritism in public procurement.
\newblock \emph{European Journal on Criminal Policy and Research}, 26\penalty0
  (4):\penalty0 411--430, 2020.

\bibitem[Della~Porta and Vannucci(2016)]{della2016hidden}
Donatella Della~Porta and Alberto Vannucci.
\newblock \emph{The hidden order of corruption: an institutional approach}.
\newblock Routledge, 2016.

\bibitem[Dominguez et~al.(2020)Dominguez, Pantoja, Pico, Mateos, del Mar
  Alonso-Almeida, and González]{dominguez2020panama}
David Dominguez, Odette Pantoja, Pablo Pico, Miguel Mateos, María del Mar
  Alonso-Almeida, and Mario González.
\newblock Panama papers' offshoring network behavior.
\newblock \emph{Heliyon}, 6\penalty0 (6), 2020.

\bibitem[Duxbury and Haynie(2018)]{duxbury2018network}
Scott~W Duxbury and Dana~L Haynie.
\newblock The network structure of opioid distribution on a darknet
  cryptomarket.
\newblock \emph{Journal of quantitative criminology}, 34\penalty0 (4):\penalty0
  921--941, 2018.

\bibitem[Elnawawy et~al.(2021)Elnawawy, Okasha, and
  Hosny]{Elnawawy2019ABM_review}
Shaymaa~M. Elnawawy, Ahmed~E. Okasha, and Hazem~A. Hosny.
\newblock Agent-based models of administrative corruption: An overview.
\newblock \emph{International Journal of Modelling and Simulation}, 2021.

\bibitem[Enikolopov et~al.(2018)Enikolopov, Petrova, and
  Sonin]{enikolopov2018social}
Ruben Enikolopov, Maria Petrova, and Konstantin Sonin.
\newblock Social media and corruption.
\newblock \emph{American Economic Journal: Applied Economics}, 10\penalty0
  (1):\penalty0 150--74, 2018.

\bibitem[Ensign et~al.(2018)Ensign, Friedler, Neville, Scheidegger, and
  Venkatasubramanian]{ensign2018runaway}
Danielle Ensign, Sorelle~A Friedler, Scott Neville, Carlos Scheidegger, and
  Suresh Venkatasubramanian.
\newblock Runaway feedback loops in predictive policing.
\newblock In \emph{Conference on Fairness, Accountability and Transparency},
  pages 160--171. PMLR, 2018.

\bibitem[Farmer and Foley(2009)]{Farmer2009Nature}
J~Doyne Farmer and Duncan~K. Foley.
\newblock The economy needs agent-based modeling.
\newblock \emph{Nature}, 460:\penalty0 685--686, 2009.

\bibitem[Fazekas and Czibik(2021)]{fazekas2021measuring}
Mih{\'a}ly Fazekas and {\'A}gnes Czibik.
\newblock Measuring regional quality of government: the public spending quality
  index based on government contracting data.
\newblock \emph{Regional Studies}, pages 1--14, 2021.

\bibitem[Fazekas and Kocsis(2020)]{fazekas2020uncovering}
Mih{\'a}ly Fazekas and G{\'a}bor Kocsis.
\newblock Uncovering high-level corruption: Cross-national objective corruption
  risk indicators using public procurement data.
\newblock \emph{British Journal of Political Science}, 50\penalty0
  (1):\penalty0 155--164, 2020.

\bibitem[Fazekas and T{\'o}th(2016)]{fazekas2016corruption}
Mih{\'a}ly Fazekas and Istv{\'a}n~J{\'a}nos T{\'o}th.
\newblock From corruption to state capture: A new analytical framework with
  empirical applications from hungary.
\newblock \emph{Political Research Quarterly}, 69\penalty0 (2):\penalty0
  320--334, 2016.

\bibitem[Fazekas and Wachs(2020)]{fazekas2020corruption}
Mih{\'a}ly Fazekas and Johannes Wachs.
\newblock Corruption and the network structure of public contracting markets
  across government change.
\newblock \emph{Politics and Governance}, 8\penalty0 (2):\penalty0 153--166,
  2020.

\bibitem[Fazekas et~al.(2021)Fazekas, Ferrali, and Wachs]{fazekas2021agency}
Mih{\'a}ly Fazekas, Romain Ferrali, and Johannes Wachs.
\newblock Agency politicisation, campaign contributions, and favouritism in us
  federal government contracting.
\newblock 2021.

\bibitem[Ferrali(2020)]{ferrali2020partners}
Romain Ferrali.
\newblock Partners in crime? corruption as a criminal network.
\newblock \emph{Games and Economic Behavior}, 124:\penalty0 319--353, 2020.

\bibitem[Ferraz and Finan(2008)]{ferraz2008exposing}
Claudio Ferraz and Frederico Finan.
\newblock Exposing corrupt politicians: the effects of brazil's publicly
  released audits on electoral outcomes.
\newblock \emph{The Quarterly journal of economics}, 123\penalty0 (2):\penalty0
  703--745, 2008.

\bibitem[Fjelde and Hegre(2014)]{fjelde2014political}
Hanne Fjelde and H{\aa}vard Hegre.
\newblock Political corruption and institutional stability.
\newblock \emph{Studies in Comparative International Development}, 49\penalty0
  (3):\penalty0 267--299, 2014.

\bibitem[Garcia-Bernardo et~al.(2017)Garcia-Bernardo, Fichtner, Takes, and
  Heemskerk]{garcia2017uncovering}
Javier Garcia-Bernardo, Jan Fichtner, Frank~W Takes, and Eelke~M Heemskerk.
\newblock Uncovering offshore financial centers: Conduits and sinks in the
  global corporate ownership network.
\newblock \emph{Scientific Reports}, 7\penalty0 (1):\penalty0 1--10, 2017.

\bibitem[Gilbert(2008)]{Gilbert2008ABM}
N.~Gilbert.
\newblock \emph{Agent-Based Models}.
\newblock London Sage, 2008.

\bibitem[Granovetter(1985)]{granovetter1985economic}
Mark Granovetter.
\newblock Economic action and social structure: The problem of embeddedness.
\newblock \emph{{American Journal of Sociology}}, 91\penalty0 (3):\penalty0
  481--510, 1985.

\bibitem[Grzymala-Busse(2008)]{grzymala2008beyond}
Anna Grzymala-Busse.
\newblock Beyond clientelism: Incumbent state capture and state formation.
\newblock \emph{Comparative political studies}, 41\penalty0 (4-5):\penalty0
  638--673, 2008.

\bibitem[Gulino and Masera(2021)]{gulino2021contagious}
Giorgio Gulino and Federico Masera.
\newblock Contagious dishonesty: Corruption scandals and supermarket theft.
\newblock Technical report, 2021.

\bibitem[Gupta(1998)]{gupta1998does}
Sanjeev Gupta.
\newblock \emph{Does corruption affect income inequality and poverty?}
\newblock International Monetary Fund, 1998.

\bibitem[Hammond(2000)]{Hammond2000Endogenous}
Ross Hammond.
\newblock Endogenous transition dynamics in corruption: An agent-based computer
  model.
\newblock Technical report, Washington (DC): Center on Social and Economic
  Dynamics, 2000.

\bibitem[Heywood(2017)]{heywood2017rethinking}
Paul~M Heywood.
\newblock Rethinking corruption: Hocus-pocus, locus and focus.
\newblock \emph{Slavonic \& East European Review}, 95\penalty0 (1):\penalty0
  21--48, 2017.

\bibitem[Heywood and Rose(2014)]{heywood2014close}
Paul~M Heywood and Jonathan Rose.
\newblock “close but no cigar”: the measurement of corruption.
\newblock \emph{Journal of Public Policy}, 34\penalty0 (3):\penalty0 507--529,
  2014.

\bibitem[Jancsics(2017)]{jancsics2017offshoring}
David Jancsics.
\newblock Offshoring at home? domestic use of shell companies for corruption.
\newblock \emph{Public integrity}, 19\penalty0 (1):\penalty0 4--21, 2017.

\bibitem[Jeude et~al.(2019)Jeude, Aste, and Caldarelli]{jeude2019multilayer}
J~A van Lidth~de Jeude, T~Aste, and G~Caldarelli.
\newblock The multilayer structure of corporate networks.
\newblock \emph{New Journal of Physics}, 21\penalty0 (2), 2019.

\bibitem[Joaristi et~al.(2019)Joaristi, Serra, and
  Spezzano]{joaristi2019social}
Mikel Joaristi, Edoardo Serra, and Francesca Spezzano.
\newblock Detecting suspicious entities in offshore leaks networks.
\newblock \emph{Social Network Analysis and Mining}, 9\penalty0 (62), 2019.

\bibitem[Jong-Sung and Khagram(2005)]{jong2005comparative}
You Jong-Sung and Sanjeev Khagram.
\newblock A comparative study of inequality and corruption.
\newblock \emph{American sociological review}, 70\penalty0 (1):\penalty0
  136--157, 2005.

\bibitem[Joseph and Smith(2021)]{joseph2021ties}
Jared Joseph and Chris~M Smith.
\newblock The ties that bribe: Corruption's embeddedness in chicago organized
  crime.
\newblock \emph{Criminology}, 2021.

\bibitem[Kejriwal and Dang(2020)]{kejriwal2020structural}
Mayank Kejriwal and Akarsh Dang.
\newblock Structural studies of the global networks exposed in the panama
  papers.
\newblock \emph{Appl Netw Sci}, 5\penalty0 (63), 2020.

\bibitem[Kert{\'e}sz and Wachs(2021)]{kertesz2021complexity}
J{\'a}nos Kert{\'e}sz and Johannes Wachs.
\newblock Complexity science approach to economic crime.
\newblock \emph{Nature Reviews Physics}, 3\penalty0 (2):\penalty0 70--71, 2021.

\bibitem[Klitgaard(1988)]{klitgaard1988controlling}
Robert Klitgaard.
\newblock \emph{Controlling corruption}.
\newblock Univ of California Press, 1988.

\bibitem[Kurniawan and Felisiano(2020)]{kurniawan2020disclosure}
Riza~Alifianto Kurniawan and Iqbal Felisiano.
\newblock Disclosure of beneficial ownership to eradicate transnational
  financial crime.
\newblock In \emph{Advancing Rule of Law in a Global Context}, pages 14--18.
  CRC Press, 2020.

\bibitem[Lum and Isaac(2016)]{lum2016predict}
Kristian Lum and William Isaac.
\newblock To predict and serve?
\newblock \emph{Significance}, 13\penalty0 (5):\penalty0 14--19, 2016.

\bibitem[Luna-Pla and Nicol{\'a}s-Carlock(2020)]{luna2020corruption}
Issa Luna-Pla and Jos{\'e}~R Nicol{\'a}s-Carlock.
\newblock Corruption and complexity: a scientific framework for the analysis of
  corruption networks.
\newblock \emph{Applied Network Science}, 5\penalty0 (1):\penalty0 1--18, 2020.

\bibitem[Lyra et~al.(2021)Lyra, Curado, Dam{\'a}sio, Ba{\c{c}}{\~a}o, and
  Pinheiro]{lyra021firmfirm}
Marcos Lyra, Ant{\'o}nio Curado, Bruno Dam{\'a}sio, Fernando Ba{\c{c}}{\~a}o,
  and Fl{\'a}vio~L Pinheiro.
\newblock Characterization of the firm–firm public procurement co-bidding
  network from the state of ceará (brazil) municipalities.
\newblock \emph{Applied Network Science}, 6\penalty0 (77), 2021.

\bibitem[MacRae(1982)]{MacRae1982Underdevelopment}
John MacRae.
\newblock Underdevelopment and the economics of corruption: A game theory
  approach.
\newblock \emph{World Development}, 10:\penalty0 677--687, 1982.

\bibitem[Mauro(1995)]{mauro1995corruption}
Paolo Mauro.
\newblock Corruption and growth.
\newblock \emph{{The Quarterly Journal of Economics}}, 110\penalty0
  (3):\penalty0 681--712, 1995.

\bibitem[Mungiu-Pippidi(2017)]{mungiu2017time}
Alina Mungiu-Pippidi.
\newblock The time has come for evidence-based anticorruption.
\newblock \emph{Nature: Human Behavior}, 2017.

\bibitem[Nizhnikau(2020)]{nizhnikau2020love}
Ryhor Nizhnikau.
\newblock Love the tender: Prozorro and anti-corruption reforms after the
  euromaidan revolution.
\newblock \emph{Problems of Post-Communism}, pages 1--14, 2020.

\bibitem[Nye(1967)]{nye1967corruption}
Joseph~S Nye.
\newblock Corruption and political development: A cost-benefit analysis.
\newblock \emph{American political science review}, 61\penalty0 (2):\penalty0
  417--427, 1967.

\bibitem[OECD.Stat(2017)]{oecdprocurement}
OECD.Stat.
\newblock Government at a glance - 2017 edition: Public procurement.
\newblock \url{https://stats.oecd.org/Index.aspx?QueryId=78413}, 2017.
\newblock Accessed: 2018-08-09.

\bibitem[Olken and Pande(2012)]{Olken2012Developing}
B.A. Olken and R.~Pande.
\newblock Corruption in developing countries.
\newblock \emph{Annual Review of Economics}, 4:\penalty0 479--505, 2012.

\bibitem[Olken(2007)]{olken2007monitoring}
Benjamin~A Olken.
\newblock Monitoring corruption: evidence from a field experiment in indonesia.
\newblock \emph{Journal of Political Economy}, 115\penalty0 (2):\penalty0
  200--249, 2007.

\bibitem[Osborne(2010)]{Osborne2010Gametheory}
M.J. Osborne.
\newblock \emph{An intorduction to game theory}.
\newblock Oxford UP, 2010.

\bibitem[Papachristos et~al.(2013)Papachristos, Hureau, and
  Braga]{papachristos2013corner}
Andrew~V Papachristos, David~M Hureau, and Anthony~A Braga.
\newblock The corner and the crew: The influence of geography and social
  networks on gang violence.
\newblock \emph{American sociological review}, 78\penalty0 (3):\penalty0
  417--447, 2013.

\bibitem[Paquet-Clouston et~al.(2019)Paquet-Clouston, Haslhofer, and
  Dupont]{paquet2019ransomware}
Masarah Paquet-Clouston, Bernhard Haslhofer, and Benoit Dupont.
\newblock Ransomware payments in the bitcoin ecosystem.
\newblock \emph{Journal of Cybersecurity}, 5\penalty0 (1):\penalty0 tyz003,
  2019.

\bibitem[Persson et~al.(2013)Persson, Rothstein, and
  Teorell]{persson2013anticorruption}
Anna Persson, Bo~Rothstein, and Jan Teorell.
\newblock Why anticorruption reforms fail—systemic corruption as a collective
  action problem.
\newblock \emph{Governance}, 26\penalty0 (3):\penalty0 449--471, 2013.

\bibitem[Persson and Tabellini(2002)]{persson2002political}
Torsten Persson and Guido~Enrico Tabellini.
\newblock \emph{Political economics: explaining economic policy}.
\newblock MIT press, {}, 2002.

\bibitem[Pirro and Della~Porta(2021)]{pirro2021corruption}
Andrea~LP Pirro and Donatella Della~Porta.
\newblock On corruption and state capture: The struggle of anti-corruption
  activism in hungary.
\newblock \emph{Europe-Asia Studies}, 73\penalty0 (3):\penalty0 433--450, 2021.

\bibitem[Railsback and Grimm(2019)]{Railsback2019ABMbook}
Steven~F. Railsback and Volker Grimm.
\newblock \emph{Agent-Based and Individual-Based Modeling: A Practical
  Introduction}.
\newblock Princeton UP, 2019.

\bibitem[Ren et~al.(2019)Ren, Gleinig, Helbing, and
  Antulov-Fantulin]{ren2019generalized}
Xiao-Long Ren, Niels Gleinig, Dirk Helbing, and Nino Antulov-Fantulin.
\newblock Generalized network dismantling.
\newblock \emph{Proceedings of the National Academy of Sciences}, 116\penalty0
  (14):\penalty0 6554--6559, 2019.

\bibitem[Ribeiro et~al.(2018)Ribeiro, Alves, Martins, Lenzi, and
  Perc]{ribeiro2018dynamical}
Haroldo~V Ribeiro, Luiz~GA Alves, Alvaro~F Martins, Ervin~K Lenzi, and
  Matja{\v{z}} Perc.
\newblock The dynamical structure of political corruption networks.
\newblock \emph{Journal of Complex Networks}, 6\penalty0 (6):\penalty0
  989--1003, 2018.

\bibitem[Rodr{\'\i}guez-Pose and Di~Cataldo(2014)]{rodriguez2014quality}
Andr{\'e}s Rodr{\'\i}guez-Pose and Marco Di~Cataldo.
\newblock Quality of government and innovative performance in the regions of
  europe.
\newblock \emph{Journal of Economic Geography}, 15\penalty0 (4):\penalty0
  673--706, 2014.

\bibitem[Rothstein(2013)]{rothstein2013corruption}
Bo~Rothstein.
\newblock Corruption and social trust: Why the fish rots from the head down.
\newblock \emph{social research}, 80\penalty0 (4):\penalty0 1009--1032, 2013.

\bibitem[Rothstein et~al.(2013)Rothstein, Charron, and
  Lapuente]{rothstein2013quality}
Bo~Rothstein, Nicholas Charron, and Victor Lapuente.
\newblock \emph{Quality of government and corruption from a European
  perspective: a comparative study on the quality of government in EU regions}.
\newblock Edward Elgar Publishing, {}, 2013.

\bibitem[Situngkir(2003)]{Situngkir2003Money-Scape}
Hokky Situngkir.
\newblock Money-scape: A generic agent-based model of corruption.
\newblock \emph{Computational Economics Archive}, 405008:\penalty0
  ~available~at https://core.ac.uk/reader/9314990, 2003.

\bibitem[Situngkir(2004)]{Situngkir2004StructuralDyn}
Hokky Situngkir.
\newblock The structural dynamics of corruption: Artificial society approach.
\newblock \emph{ArXiv}, page ~available~at
  https://arxiv.org/ftp/nlin/papers/0403/0403042.pdf, 2004.

\bibitem[Soylu et~al.(2019)Soylu, Elves{\ae}ter, Turk, Roman, Corcho, Simperl,
  Makgill, Taggart, Grobelnik, and Lech]{soylu2019overview}
Ahmet Soylu, Brian Elves{\ae}ter, Philip Turk, Dumitru Roman, {\'O}scar Corcho,
  Elena Simperl, Ian Makgill, Chris Taggart, Marko Grobelnik, and
  Till~Christopher Lech.
\newblock An overview of the tbfy knowledge graph for public procurement.
\newblock In \emph{ISWC Satellites}, pages 53--56, 2019.

\bibitem[Szabo and Fath(2007)]{Szabo2007Evolutionary}
Gyorgy Szabo and Gabor Fath.
\newblock Evolutionary games on graphs.
\newblock \emph{Physics Reports}, 446:\penalty0 97--216, 2007.

\bibitem[Torsello and Venard(2016)]{torsello2016anthropology}
Davide Torsello and Bertrand Venard.
\newblock The anthropology of corruption.
\newblock \emph{Journal of Management Inquiry}, 25\penalty0 (1):\penalty0
  34--54, 2016.

\bibitem[{Transparency International}(2007)]{transparency2007global}
{Transparency International}.
\newblock Global corruption report 2007.
\newblock Technical report, 2007.

\bibitem[Wachs et~al.(2019)Wachs, Yasseri, Lengyel, and
  Kert{\'e}sz]{wachs2019social}
Johannes Wachs, Taha Yasseri, Bal{\'a}zs Lengyel, and J{\'a}nos Kert{\'e}sz.
\newblock Social capital predicts corruption risk in towns.
\newblock \emph{Royal Society open science}, 6\penalty0 (4):\penalty0 182103,
  2019.

\bibitem[Wachs et~al.(2021)Wachs, Fazekas, and
  Kert{\'e}sz]{wachs2021corruption}
Johannes Wachs, Mih{\'a}ly Fazekas, and J{\'a}nos Kert{\'e}sz.
\newblock Corruption risk in contracting markets: a network science
  perspective.
\newblock \emph{International Journal of Data Science and Analytics},
  12\penalty0 (1):\penalty0 45--60, 2021.

\bibitem[Weisburd et~al.(2017)Weisburd, Braga, Groff, and
  Wooditch]{weisburd2017can}
David Weisburd, Anthony~A Braga, Elizabeth~R Groff, and Alese Wooditch.
\newblock Can hot spots policing reduce crime in urban areas? an agent-based
  simulation.
\newblock \emph{Criminology}, 55\penalty0 (1):\penalty0 137--173, 2017.

\bibitem[Weisel and Shalvi(2015)]{weisel2015collaborative}
Ori Weisel and Shaul Shalvi.
\newblock The collaborative roots of corruption.
\newblock \emph{Proceedings of the National Academy of Sciences}, 112\penalty0
  (34):\penalty0 10651--10656, 2015.

\bibitem[Witko(2011)]{witko2011campaign}
Christopher Witko.
\newblock Campaign contributions, access, and government contracting.
\newblock \emph{Journal of Public Administration Research and Theory},
  21\penalty0 (4):\penalty0 761--778, 2011.

\bibitem[Wood et~al.(2019)Wood, Roithmayr, and Papachristos]{wood2019network}
George Wood, Daria Roithmayr, and Andrew~V Papachristos.
\newblock The network structure of police misconduct.
\newblock \emph{Socius}, 5:\penalty0 2378023119879798, 2019.

\bibitem[Zausinova et~al.(2020)Zausinova, Zoricak, Volosin, and
  Gazda]{Zausinova2020Complexity}
A.~Zausinova, M.~Zoricak, M.~Volosin, and V.~Gazda.
\newblock Aspects of complexity in citizen–bureaucrat corruption: an
  agent-based simulation model.
\newblock \emph{J. Econ. Interact. Coord.}, 15:\penalty0 527--552, 2020.

\end{thebibliography}

\end{document}